\documentclass[12pt]{article}
\usepackage{graphicx}

\newcommand\pubnumber{}
\newcommand\pubdate{\today}

\def\Title#1{\begin{center} {\Large #1 } \end{center}}
\def\Author#1{\begin{center}{ \sc #1} \end{center}}
\def\Address#1{\begin{center}{ \it #1} \end{center}}

\newcommand\pubblock{\rightline{\begin{tabular}{l} \pubnumber\\
         \pubdate  \end{tabular}}}
\newenvironment{Abstract}{\begin{quotation}  }{\end{quotation}}
\newenvironment{Presented}{\begin{quotation} \begin{center} 
             PRESENTED AT\end{center}\bigskip 
      \begin{center}\begin{large}}{\end{large}\end{center} \end{quotation}}
\def\Acknowledgements{\bigskip  \bigskip \begin{center} \begin{large}
             \bf ACKNOWLEDGEMENTS \end{large}\end{center}}
\def\beq{\begin{equation}}
\def\eeq#1{\label{#1}\end{equation}}
\def\eeqn{\end{equation}}

\def\beqa{\begin{eqnarray}}
\def\eeqa#1{\label{#1}\end{eqnarray}}
\def\eeqan{\end{eqnarray}}

\let\bar=\overbar

\def\Dslash{\not{\hbox{\kern-4pt $D$}}}
\def\dslash{\not{\hbox{\kern-2pt $\del$}}}

\def\BR{\mbox{\rm BR}}

\def\msb{{\bar{\ssstyle M \kern -1pt S}}}

\begin{document}
\begin{titlepage}
\pubblock

\vfill
\Title{Gain Stabilization of SiPMs\footnote{Work supported by EC as INFRA
project no. 262025, AIDA}}
\vfill
\Author{Jaroslav Cvach$^b$\footnote{Work supported by the Ministry of Education, Youth and Sports, Czech Republic, project LG130131}, {\bf  Gerald Eigen$^a$}\footnote{Work supported by the Norwegian Research Council},  Jiri Kvasnicka$^b$, Ivo Polak$^b$, Erik van der Kraiij$^a$ and Justas Zalieckas$^a$  } 
%\support}
\Address{$^a$Department of Physics,
University of Bergen, N-5007 Bergen, Norway\\
$^b$Institute of Physics (FZU), Prague, Czech Republic }

\vfill
\begin{Abstract}
The gain of SiPMs depends both on bias voltage and on temperature. For stable operations, both need to be kept constant. In an ILC calorimeter with millions of channels, this is a challenging task. It is, therefore, desirable to compensate for temperature variations by automatically readjusting the bias voltage. We have designed a  bias voltage regulator board to achieve this task. We anticipate an uncertainty on the gain stability at the level of $< 1\%$. First, we present measurements of the gain dependence on temperature and bias voltage for several SiPMs from three different manufacturers and determine their dV/dT dependence. Next, we demonstrate the performance of the gain stability with the bias voltage regulator test board on four SiPMs. 
\end{Abstract}
\vfill
\begin{Presented}
International Workshop on Future Linear Colliders, LCWS13\\
Tokyo, Japan, 11--15 November 2013
\end{Presented}
\vfill
\end{titlepage}
\def\thefootnote{\fnsymbol{footnote}}
\setcounter{footnote}{0}

\section{Introduction}

The analog hadron calorimeter (AHCAL) is one possible hadron calorimeter option for a detector operating at the international linear collider. It consists of a 30-layer steel plate scintillator tile array~\cite{hcal}. The scintillator tiles are read out with SiPMs~\cite{sipma, sipmb, sipmc} whose gain depends on both bias voltage $(V)$ and on temperature $(T)$. With increasing temperature, the gain decreases while with increasing bias voltage the gain increases. For stable operations, we need to keep the gain constant. In test beam measurements of the AHCAL physics prototype, we have corrected the gain for temperature variations offline. In a real detector with millions of channels, this is not a viable option. However, we can correct for gain changes induced by temperature variations by readjusting the bias voltage. Thus, our goal is to build a bias voltage regulator that stabilizes the gain at a level of $>99 \%$.
First, we need to measure the gain dependence versus temperature and bias voltage to extract dV/dT. 
Next, we build a bias voltage regulator test board that provides the appropriate bias voltage readjustment when the temperature
changes. This work is conducted in the framework of the EU project AIDA~\cite{AIDA}. 

\section{Test Setup}

\begin{figure}[h]
\centering
\vskip -0.4cm
\includegraphics[width=75mm]{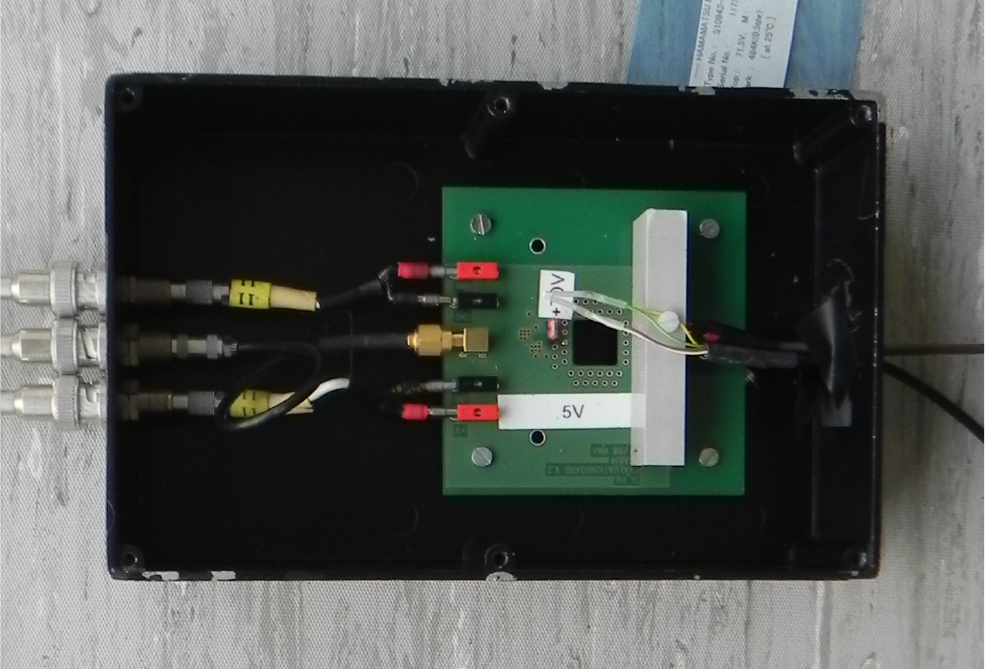}
\includegraphics[width=76mm]{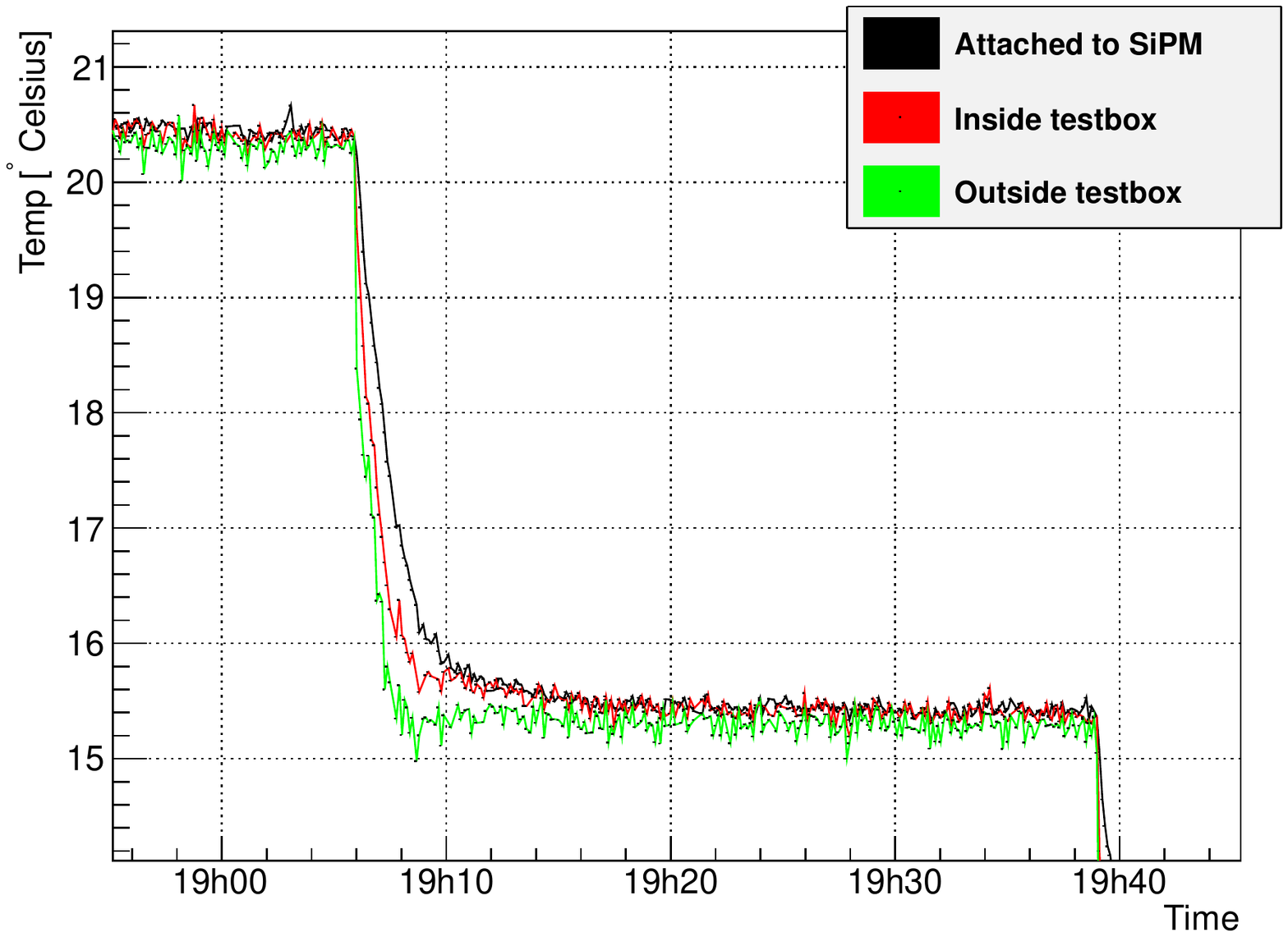}
\caption{Test box housing the SiPM, preamplifier, blue LED and temperature sensors (left). Temperature profiles of the pt~1000 sensors in the climate chamber (right) after reducing the preset temperature by $5~^\circ  \rm C$ for sensors near the SiPM (black), inside the test box (red) and in the climate chamber outside the test box (green).}
 \label{fig:temp}
\end{figure}

We perform all measurements at CERN in a climate chamber that is accurate to $0.2~^\circ   \rm C$. The SiPM is mounted inside a black metal box that also houses a preamplifier, a blue LED and temperature sensors. The preamplifier provides a gain of 8.9. The SiPM signal is recorded with a digital oscilloscope that is read out by a PC. This readout scheme was used in the T3B test beam studies~\cite{soldier, wueste}. We shine 2~ns long LED pulses on the SiPM. We measure the temperature with three pt~1000 sensors, one close to the SiPM, another one in the black box and a third one in the climate chamber. In addition, an LM35 sensor placed near the SiPM is used by the bias voltage regulator for performing the bias voltage readjustments. We vary T from $5~^\circ  \rm C$ to $40~^\circ  \rm C$ in steps of $5~^\circ  \rm C$, except for the $20^\circ-30~^\circ  \rm C$ range where we use steps of $2~^\circ  \rm C$. We take 50,000 events per run at a fixed temperature. The temperature at the SiPM and the preset temperature differ by an offset of $0.4~^\circ  \rm C$, which remains unchanged over the entire $T$ range. Figure~\ref{fig:temp}  shows temperature profiles of the pt~1000 temperature sensors after reducing the preset temperature by $5~^\circ   \rm C$. We start a new run once the temperature reaches equilibrium, which typically takes about 15 min. 

We have measured the $dG/dT$ dependence versus $V$ and the  $dG/dV$ dependence versus $T$ for 15 SiPMs from  three manufacturers (Hamamatsu, KETEK and CPTA). Table~\ref{tab:SiPM} lists their properties. The CPTA SiPMs were attached to a  $\rm 3~cm \times 3~cm$ scintillator tile. While all Hamamatsu and KETEK detectors could be illuminated directly by the blue LED, we tried to shine the LED light onto the CPTA SIPMs as closely as possible.
We selected four of them (CPTA 857, CPTA 1677, KETEK W 12 and Hamamatsu 11759) to demonstrate the performance of the bias voltage readjustment with the bias voltage regulator test board.

\begin{table}[tb]
\begin{center}
\begin{tabular}{| l|c|c|c|c|c|c|}
\hline
\textbf{Manufacturer} & \textbf{Sensitive} &  \textbf{Pixel} &  \textbf{$\#$} &
\textbf{Nominal} & \textbf{Typical} &  \textbf{Serial $\#$} \\
\textbf{and Type $\#$} & \textbf{area} & \textbf{pitch} & \textbf{pixels}&\textbf{$V_{bias}$} &\textbf{G}  & \\
	&\textbf{$\rm [mm^2]$} &  \textbf{$[ \rm \mu m ]$} &  & \textbf{ [V]} &
\textbf{$ \times [10^5] $} &
 \\ [0.7ex]
 \hline
Hamamatsu & & & & & &
\\
S10943-8584(X) & $1 \times 1$ & $50$ & 400 & 71.69 & 7.49 & 11759
\\
S10943-8584(X) & $1 \times 1$ & $50$ & 400 & 71.57 & 7.49 & 11766
\\
S10943-8584(X) & $1 \times 1$ & $50$ & 400 & 71.50 & 7.48 & 11770
\\
S10943-8584(X) & $1 \times 1$ & $50$ & 400 & 71.33 & 7.48 & 11771
\\
Sample A  & $1 \times 1$ & $20$ & 2500 & $66.7$ & $2.3$ & A1
\\
Sample B  & $1 \times 1$ & $20$ & 2500 & $73.3$ & $2.3$ & B1
\\
Sample A  & $1 \times 1$ & $15$ & 4440 & $67.2$ & $2.0$ & A2
\\
Sample B  & $1 \times 1$ & $15$ & 4440 & $74.0$ & $2.0$ & B2
\\ [0.7ex]
 \hline
 %
 % The pixel size for CPTA and pixel nrs comes from the talk by Danilov
 %  http://ilcagenda.linearcollider.org/getFile.py/access?contribId=34&sessionId=14&resId=1&materialId=slides&confId=4776
 % which, according to an email from Katja (27/05/'13), is the CPTA we have.
 %
 % But why is then the 'pixn' in the excel sheet I got from Mathias Reinecke
 % at 900-1000??
 %
 
CPTA  & & & & & &
\\
& $1 \times 1$ & $40$ & 796 & 33.4 & 7.1 & 857
\\
& $1 \times 1$ & $40$ & 796 & 33.1 & 6.3 & 922
\\
& $1 \times 1$ & $40$ & 796 & 33.3 & 6.3 & 975
\\
& $1 \times 1$ & $40$ & 796 & 33.1 & 7.0 & 1065
\\
& $1 \times 1$ & $40$ & 796 & 33.3 & 14.6 & 1677
\\ [0.7ex]
\hline
KETEK  & & & & & &
\\
MP15 V6  & $2 \times (1.2 \times 1.2) $ & 15 & 4384 & $ 28$ & $3.0$ & W8
\\
MP20 V4  & $3 \times 3$ & 20 & 12100 & $ 28$ &$6.0$  & W12
\\
\hline

 \end{tabular}
\end{center}
\caption{Properties of all tested SiPMs. The serial numbers for the CPTA
SiPMs correspond to those of the AHCAL tiles. For Hamamatsu (CPTA, KETEK) SiPMs,
operating voltage and gain are specified for a temperature of 25
(22)$~^\circ  \rm C$.}
\label{tab:SiPM}
\end{table}

\section{Layout of the Voltage Regulator Test Board}

\begin{figure}[h]
\centering
\vskip -0.cm
\includegraphics[width=90mm]{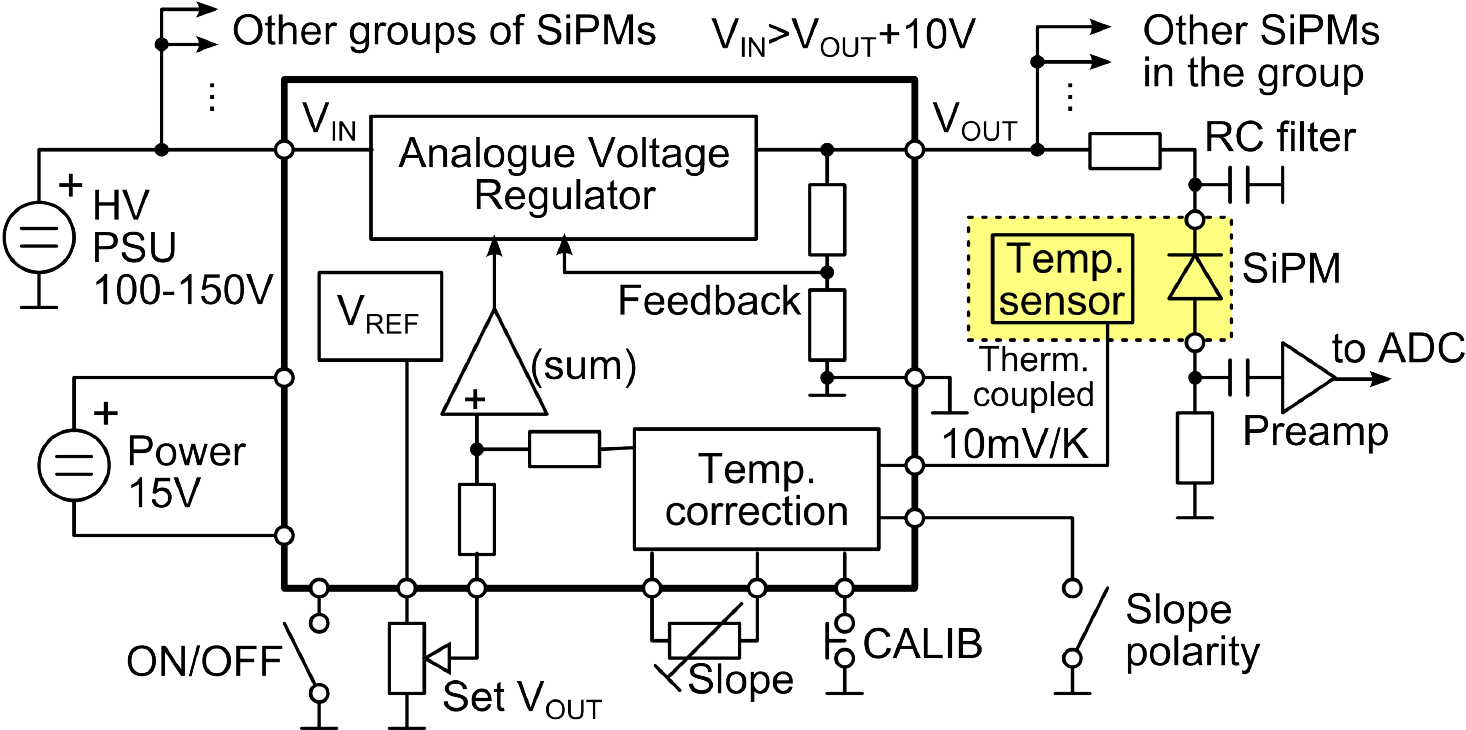}
\includegraphics[width=58mm]{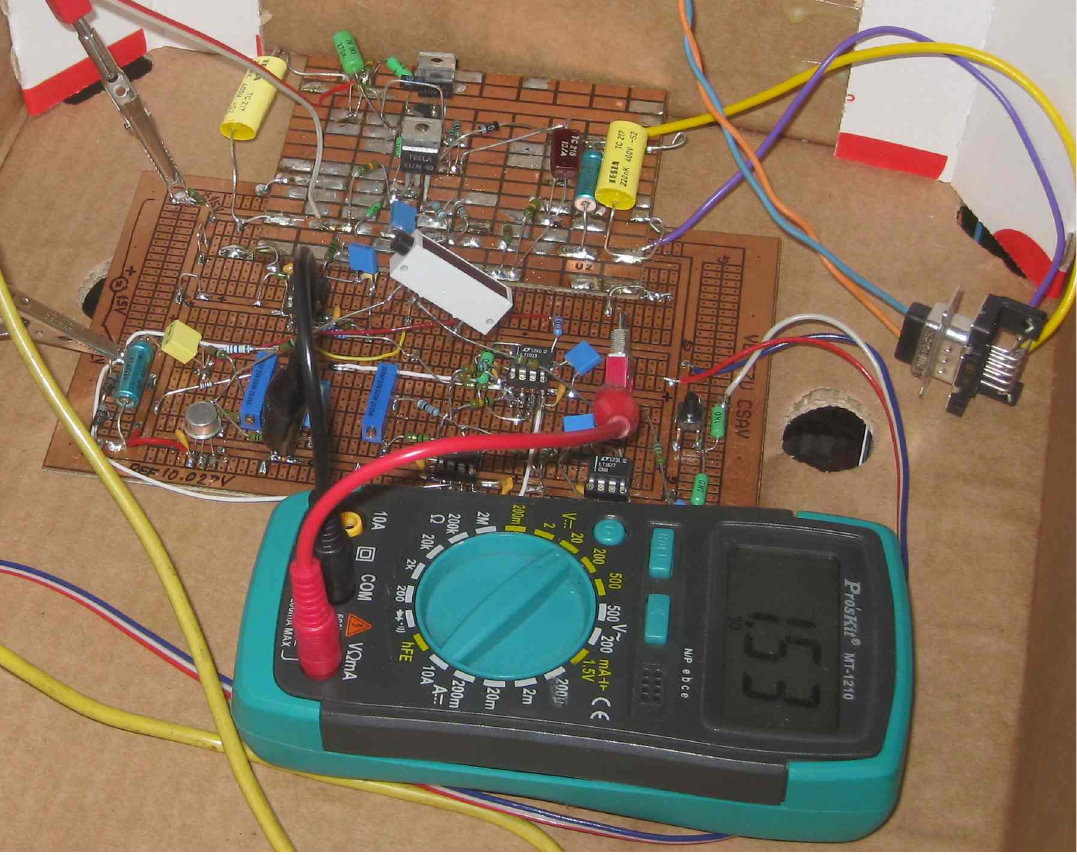}
\caption{Schematic layout of the bias voltage regulator board (left) and a photo of the first test board (right).}
 \label{fig:aps}
\end{figure}

Figure~\ref{fig:aps} (left) shows the schematic layout  of the bias voltage regulator test board and Fig.~\ref{fig:aps} (right) shows a photograph of it. The bias voltage regulator is designed to operate in the $\rm 10~V$ to $\rm 80~V$ region plus temperature correction effects. The temperature slope is variable from $\rm <1~mV/~^\circ  C$ to $\rm 100~mV/^\circ C$, both for positive and negative values. 

\begin{figure}[h]
\centering
\vskip -0.0cm
\includegraphics[width=95mm]{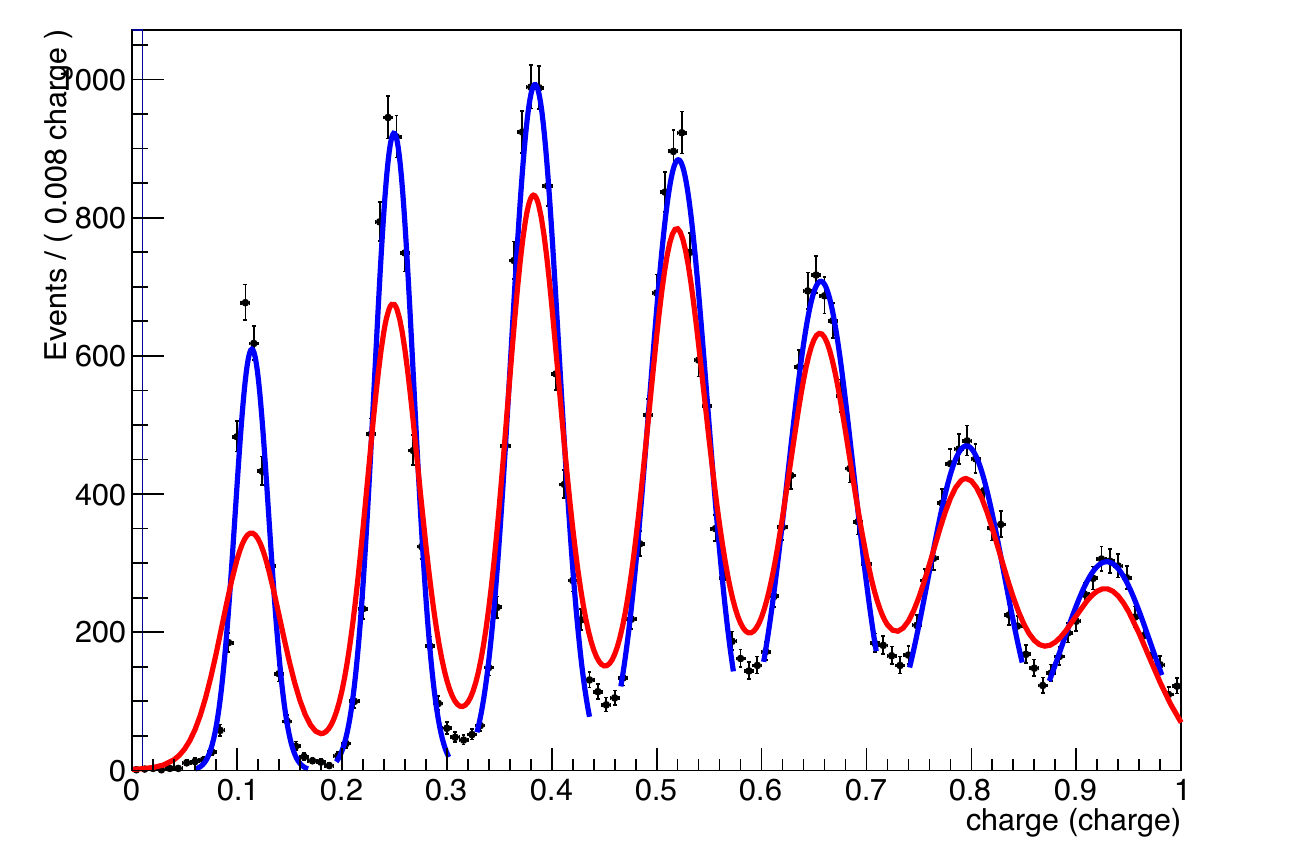}
\caption{Spectrum of single photoelectron peaks from a Hamamatsu MPPC illuminated with a blue LED. The dark (blue) curves show Gaussian fits to individual photoelectron peaks while the grey (red) curve shows a fit to a multi Gaussian kernel function.}
 \label{fig:gain}
\end{figure}

\section{Gain Determination}

\begin{figure}[h]
\centering
\vskip -0.cm
\includegraphics[width=75mm]{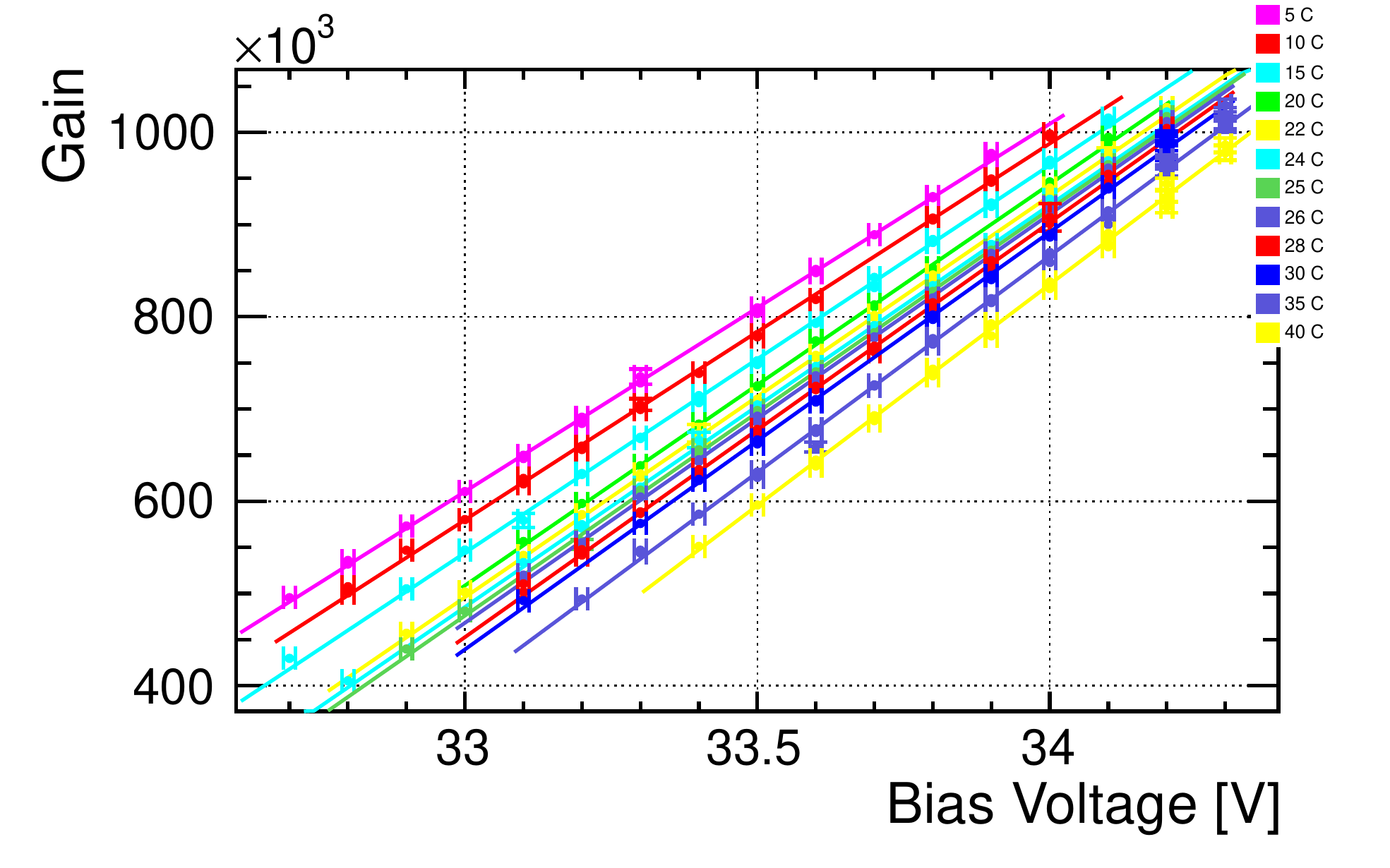}
\includegraphics[width=75mm]{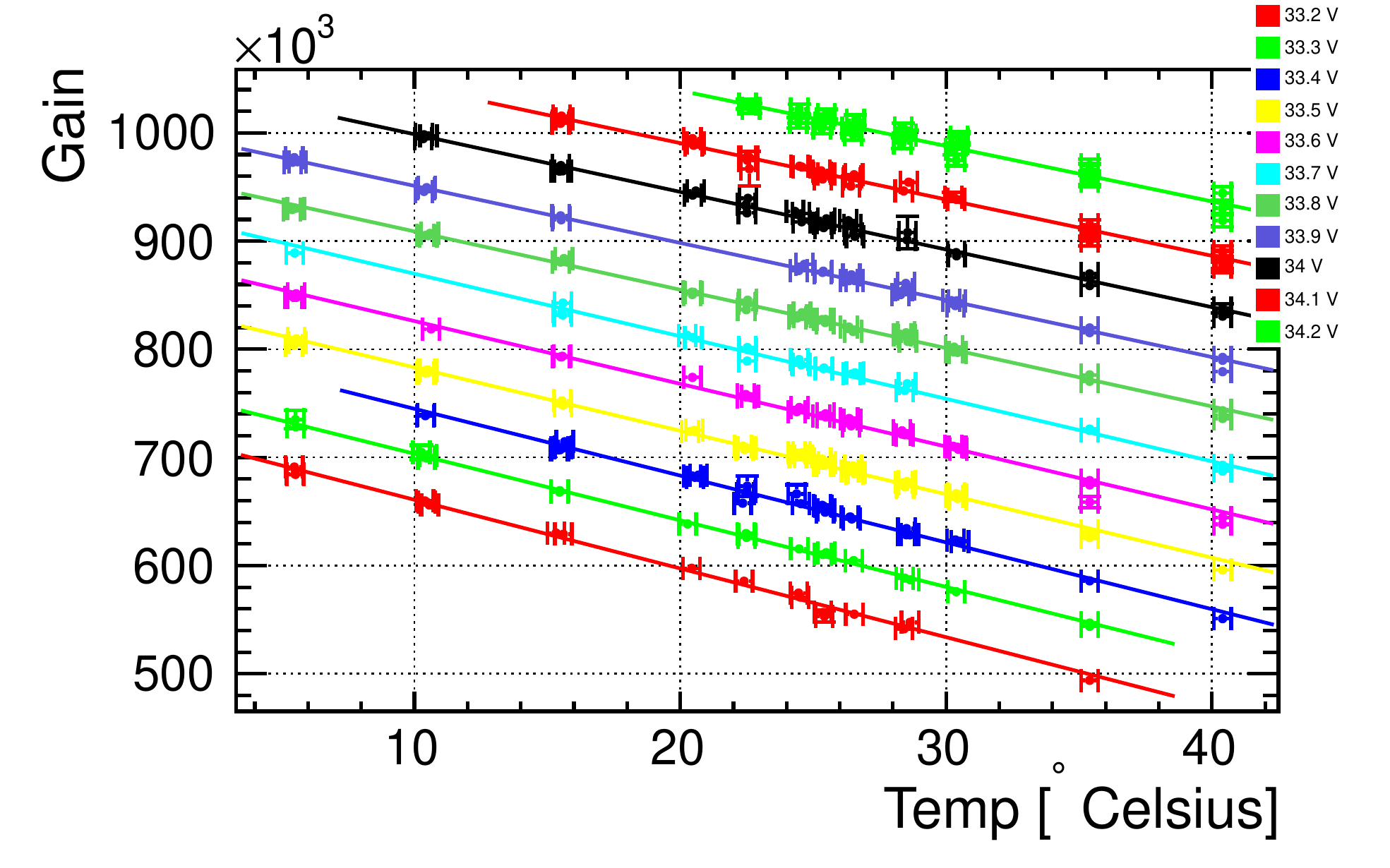}
\caption{The gain dependence on bias voltage for different temperatures between $5~^\circ   \rm C$ (magenta) and $40~^\circ   \rm C$ (yellow) for SiPM CPTA 857 (left). The gain dependence on temperature for different bias voltages from 34.2~V (green) to 33.2~V (red) for the same SiPM (right). }
 \label{fig:gv}
\end{figure}

\begin{figure}[h]
\centering
\vskip -0.cm
\includegraphics[width=49mm]{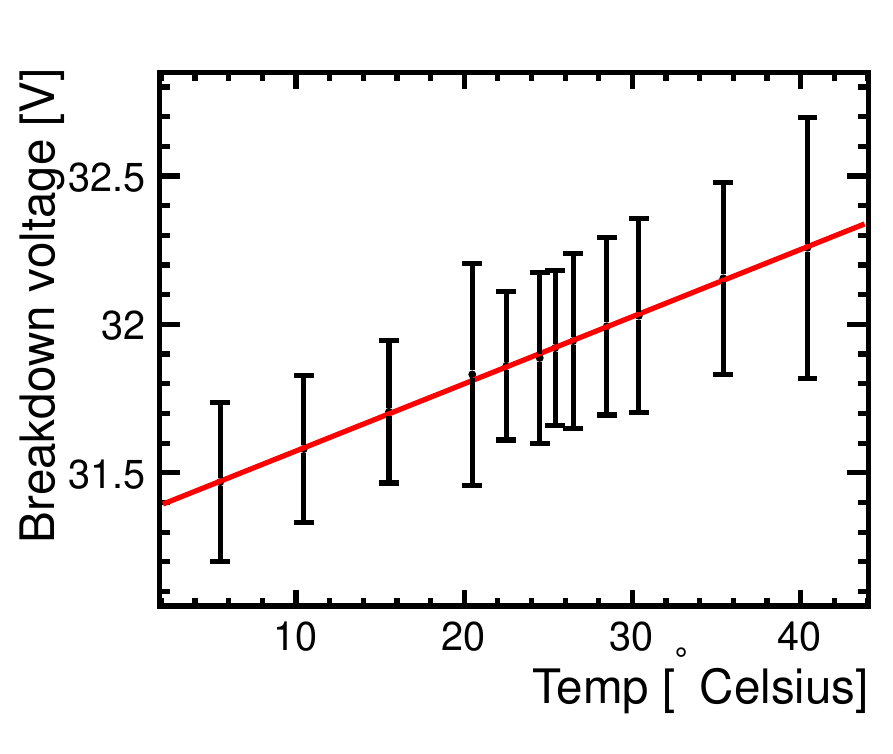}
\includegraphics[width=49mm]{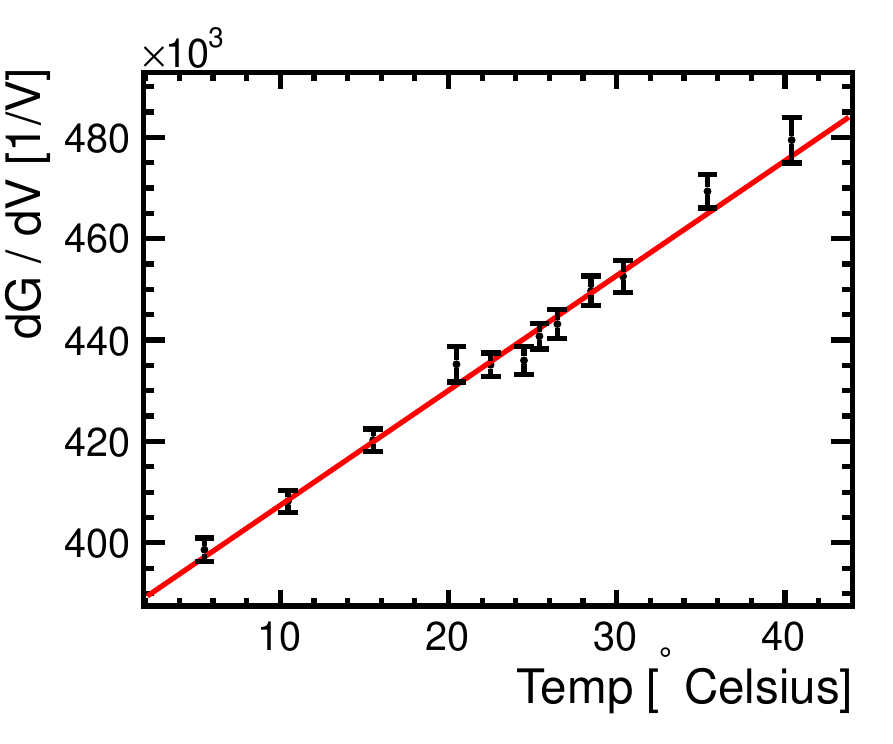}
\includegraphics[width=50mm]{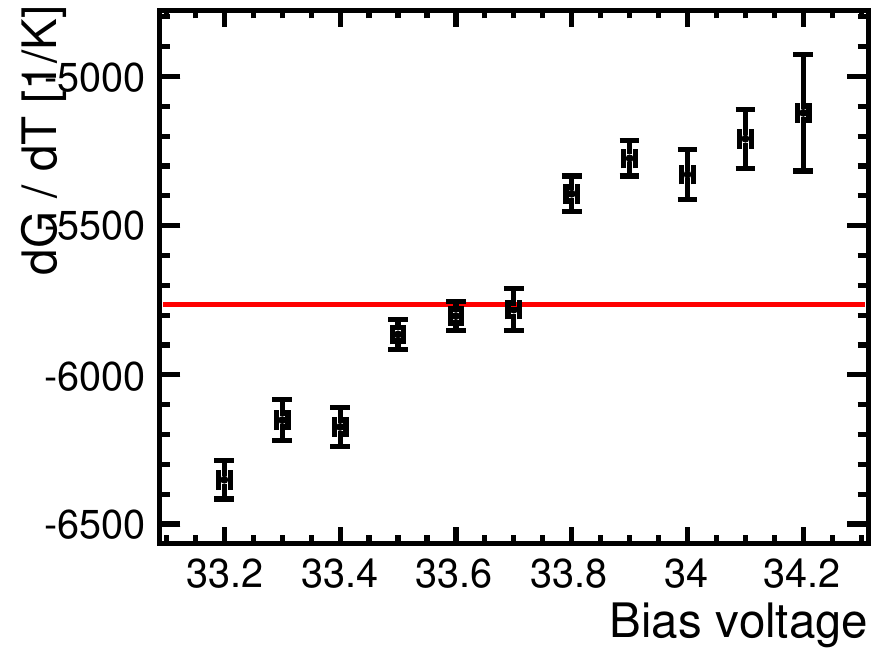}
\caption{ Breakdown voltage versus $T$ (left), $dG/dT$ versus T (middle) and  $dG/dV$ versus V (right) measured with SiPM CPTA 857.}
 \label{fig:break}
\end{figure}

Figure~\ref{fig:gain} shows a spectrum of single photoelectron peaks recorded with the Hamamatsu MPPC 11759. The red curve represents a Gaussian multi-kernel function and the blue curves show Gaussian fits of the individual photoelectron peaks. We accept only runs for which the results of the two fit methods agree within $20\%$. The gain is the distance between the pedestal peak and the first photoelectron peak. The distance between successive photoelectron peaks should be same. However, we noticed that the distance between first photoelectron peak and pedestal is slightly smaller than that between first and second photoelectron peaks. Since individual photoelectron peaks are clearly visible in Hamamatsu MPPCs, we define the gain here as the difference between the first and second photoelectron peaks. For the KTEK and CPTA detectors, the second photoelectron peak is often not clearly visible.
Therefore, we define the gain for these SiPMs as the distance between the pedestal and the first photoelectron peak. The error on the gain is determined from the uncertainties of the two fitted Gaussian peak positions added in quadrature.

For each SiPM, we measure the gain dependence on bias voltage at a fixed temperature. Each data point is determined from 50,000 80~ns long waveforms. Figure~\ref{fig:gv} (left) shows the measurements of the gain versus bias voltage for SiPM CPTA 857. For each temperature, we fit the measured points with a first-order polynomial to extract the breakdown voltage  and the slope $dG/dV$.  Figure~\ref{fig:break} (left) depicts the dependence of breakdown voltage versus temperature. We observe a linear increase of the breakdown voltage with temperature. For example at $T=25~^\circ \rm C$ and $V=33.4~\rm V$, we measure a gain of $6.4 \times 10^5$ and a breakdown voltage of $V_{break}=31.9~ \rm V.$
The slope $dG/dV$ is proportional to the capacitance of the SiPM. Figure~\ref{fig:break} (middle) depicts  the $dG/dV$ dependence on temperature. We see that the detector capacitance rises with temperature. This effect is clearly visible for most of the 15 SiPMs tested and has been observed in other studies~ \cite{Dinu}. A possible explanation is that the depletion zone in the high-gain region decreases with $T$, which in turn leads to an increase of the capacitance. We need to perform more tests to check this hypothesis. 

We fit the temperature dependence of $dG/dV$ with a zeroth-order polynomial and a first-order polynomial. 
The zeroth-order polynomial yields $\langle dG/dV \rangle= (4.32 \pm 0.01_{stat}) \times 10^5 \rm /V$ while a linear fits yields an offset of $\langle dG/dV \rangle= (4.41 \pm 0.11_{stat}) \times 10^5 \rm /V$. Since these values are in perfect agreement, we use the result from the zeroth-order polynomial fit to extract $dV/dT$.

\begin{figure}[h]
\centering
\vskip -0.0cm
\includegraphics[width=70mm]{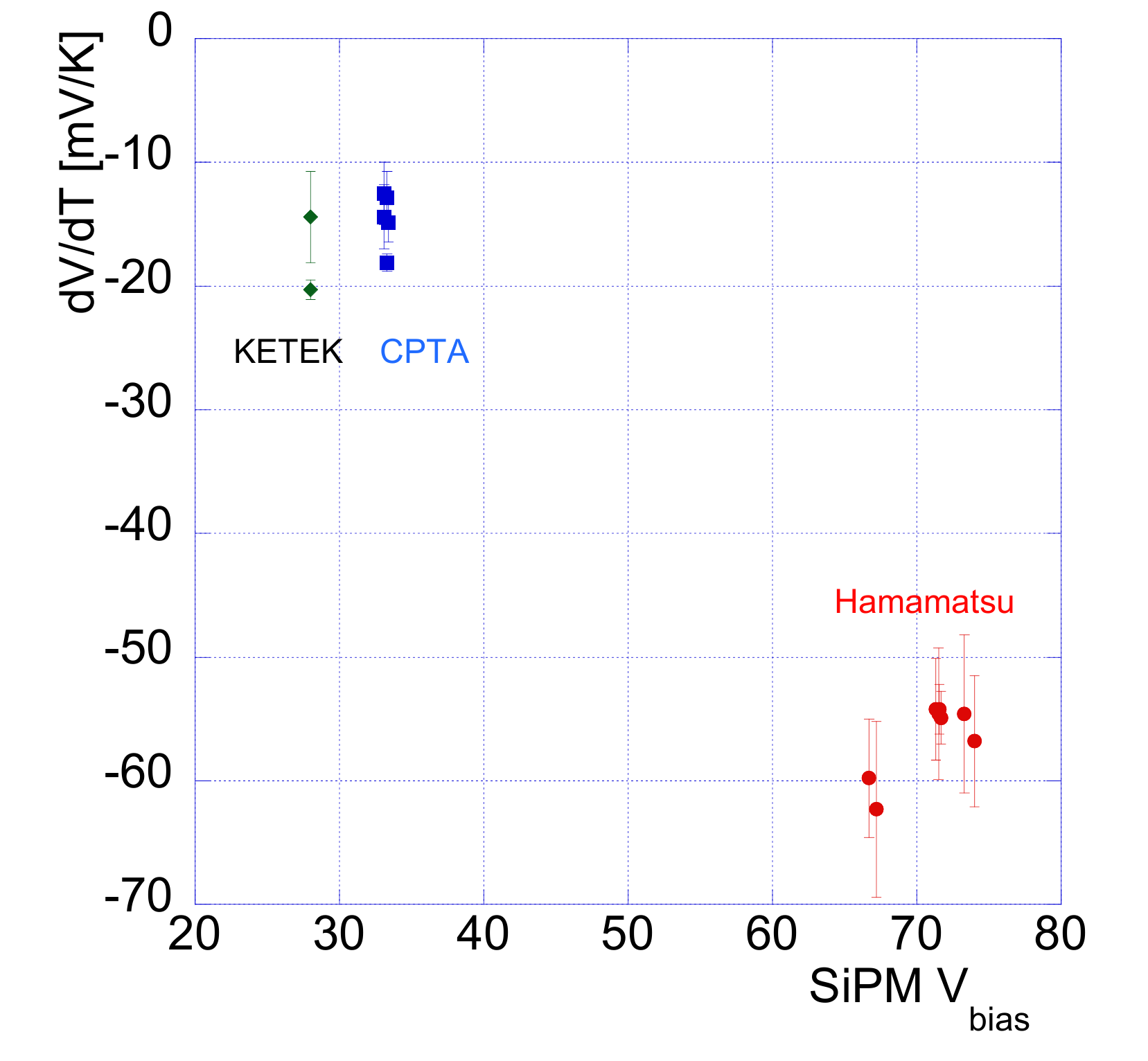}
\includegraphics[width=80mm]{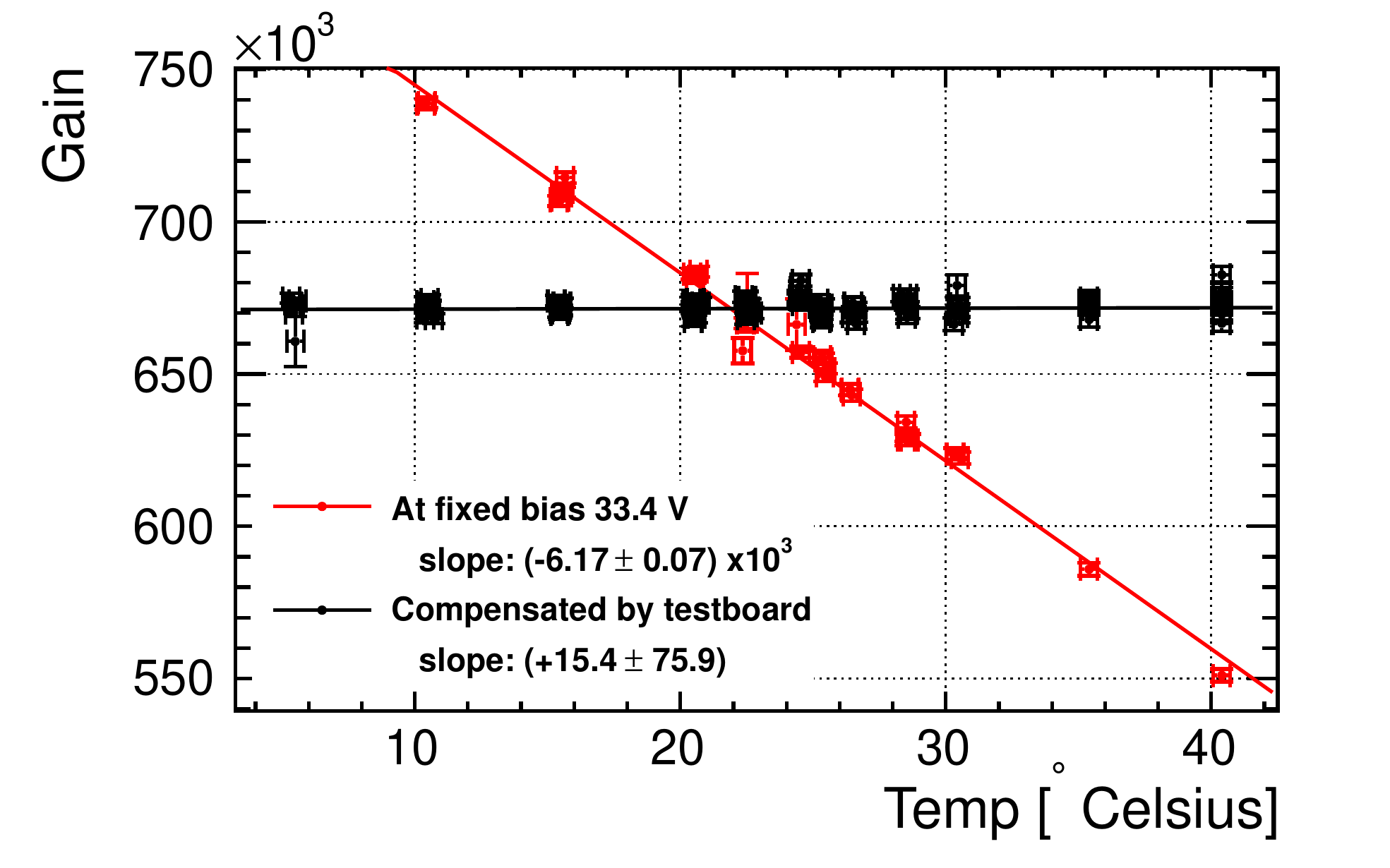}
\caption{Measured $dV/dT$ versus $V$ for all tested SiPMs (left). Gain versus temperature dependence before and after gain stabilization for SiPM CPTA 857 (right).}
 \label{fig:dvdt}
\end{figure}
  
Figure~\ref{fig:gv} (right) shows the gain dependence on temperature for fixed bias voltage for SiPM CPTA 857.  Again, we fit the data points with a first-order polynomial to extract the slope $dG/dT$. Figure~\ref{fig:break} (right) shows the $dG/dT$ dependence versus bias voltage. Again, we observe a linear dependence. A zeroth-order polynomial fit yields $\langle dG/dT \rangle= (-5.76 \pm 0.02_{stat}) \times 10^3 \rm /^\circ C$. Since this is in good agreement with the result of a first-oder polynomial fit, we use the result of the zeroth-order polynomial fit for the determination of $dV/dT$. For most of the detectors studied, we see a linear dependence of $dG/dT$ versus V.
From the  $\langle dG/dT \rangle$ and  $\langle dG/dV \rangle$ averages, we determine  $\langle dV/dT \rangle = -13.33 \pm 0.35~ \rm mV/^\circ  C$ for SiPM CPTA 857.  
Since $dG/dV$ increases with $T$ and $|dG/dT|$ decreases with $V$, the resulting $dV/dT$ values at high T and low V or low T and high V yield similar $dV/dT$ ratios leading to a $<3\%$ uncertainty on $dV/dT$. 
We define the uncertainty on the gain stability by 
\begin{equation}
 \sigma_{\Delta G/G}= \sigma_{dV/dT}\cdot \frac{1}{G} \cdot dG/dV \cdot \Delta T.
\end{equation} 
\noindent 
For SiPM CPTA 857,  $\sigma_{\Delta G/G}=0.12$. Figure~\ref{fig:dvdt} (left) shows the $dV/dT$ measurements versus bias voltage for all 15 SiPMs tested. They fall into two groups. Hamamatsu MPPCs have $dV/dT$ values in the $55 -60 \rm ~mV/^\circ C$ range and operate at higher bias voltage (65-75~V)  while KETEK and CPTA SiPMs have $dV/dT$ values around $15 - 20 \rm ~mV/^\circ C$ operating at a bias voltage of $\simeq 28~\rm V$ and $\simeq 33~\rm V$, respectively. Thus, we expect the gain stabilization to work better for KETEK and CPTA SiPMs than for Hamamatsu MPPCs.

\section{Gain Stabilization}

\begin{figure}[h]
\centering
\vskip -0.0cm
\includegraphics[width=75mm]{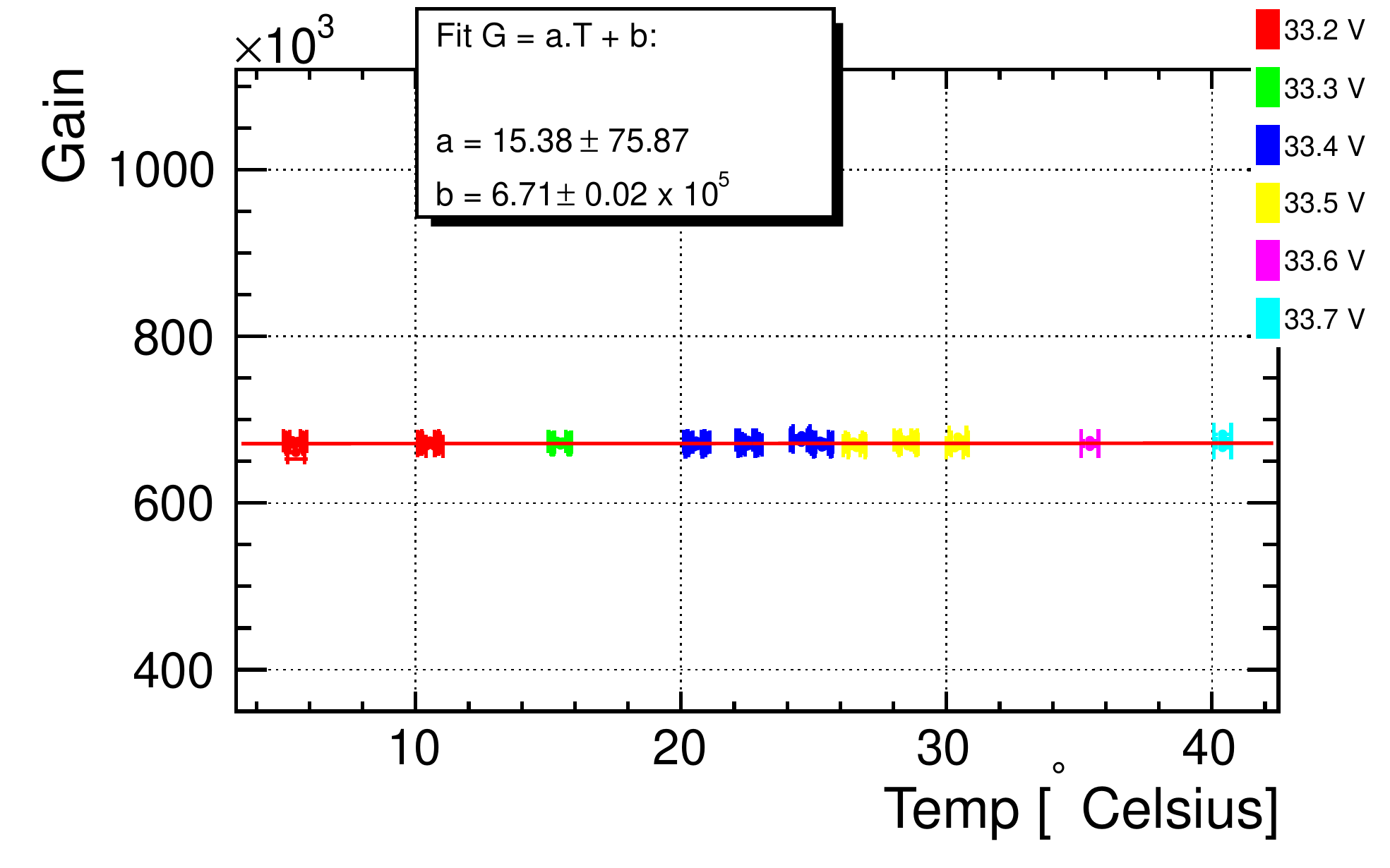}
\includegraphics[width=75mm]{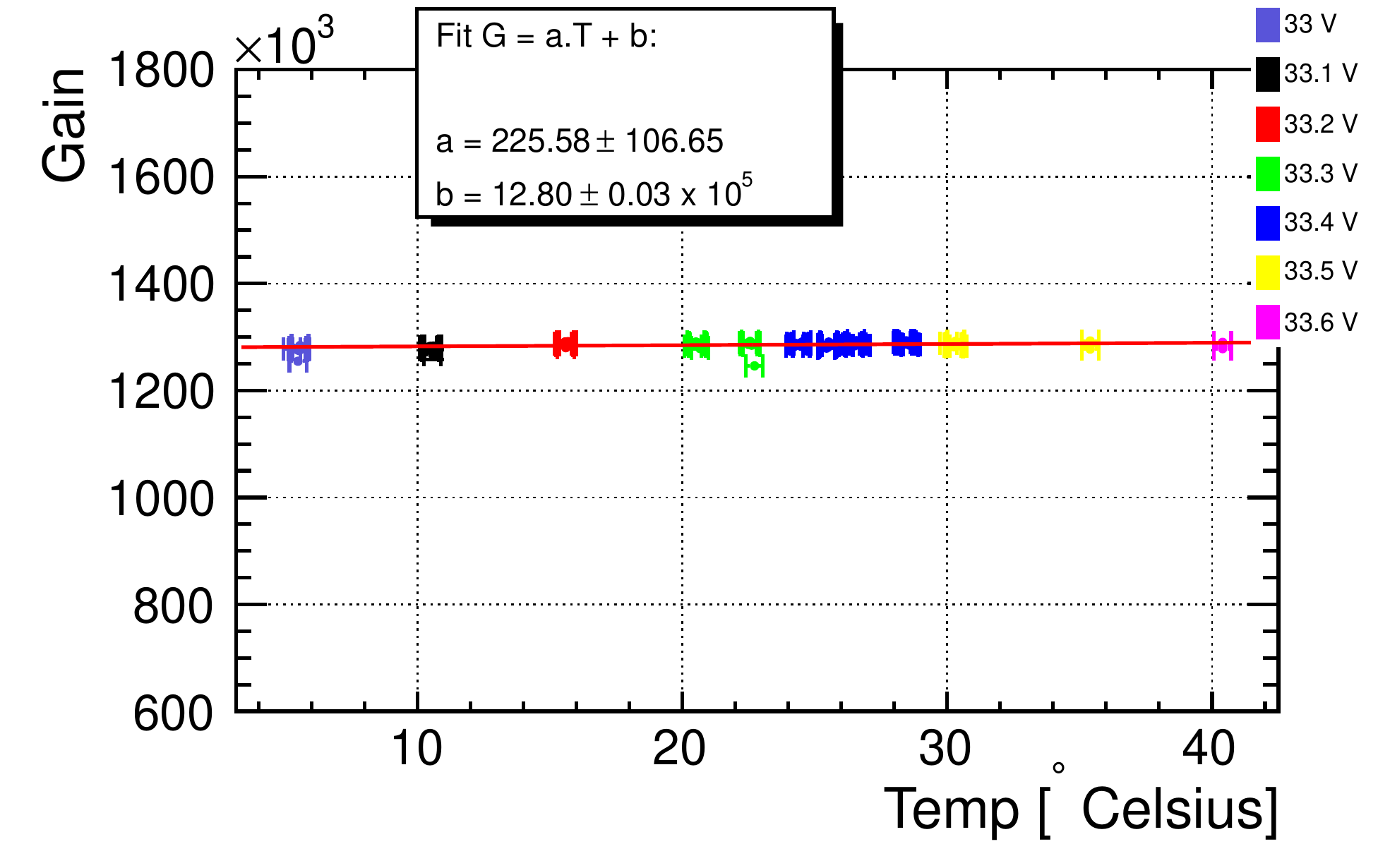}
\includegraphics[width=75mm]{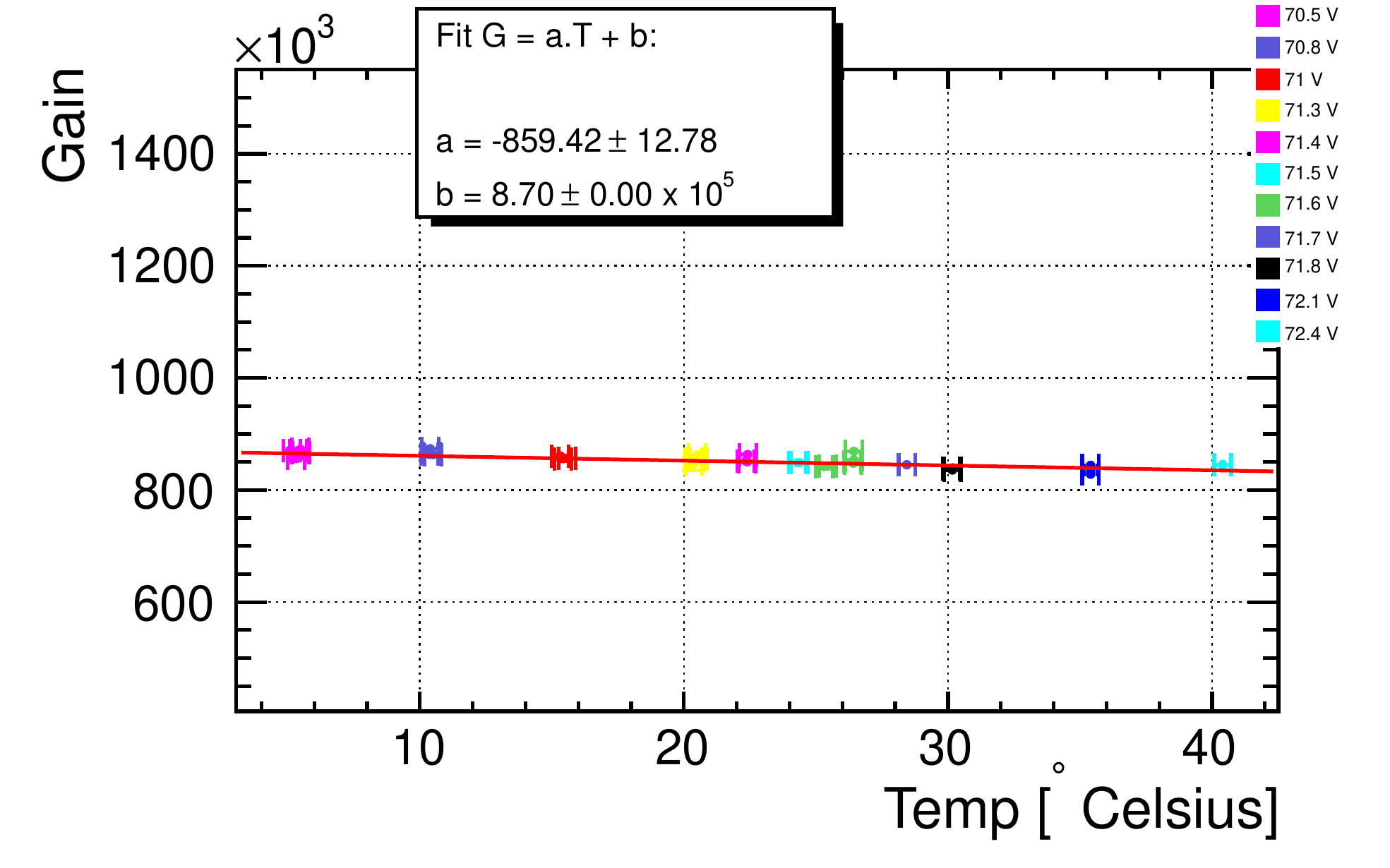}
\includegraphics[width=75mm]{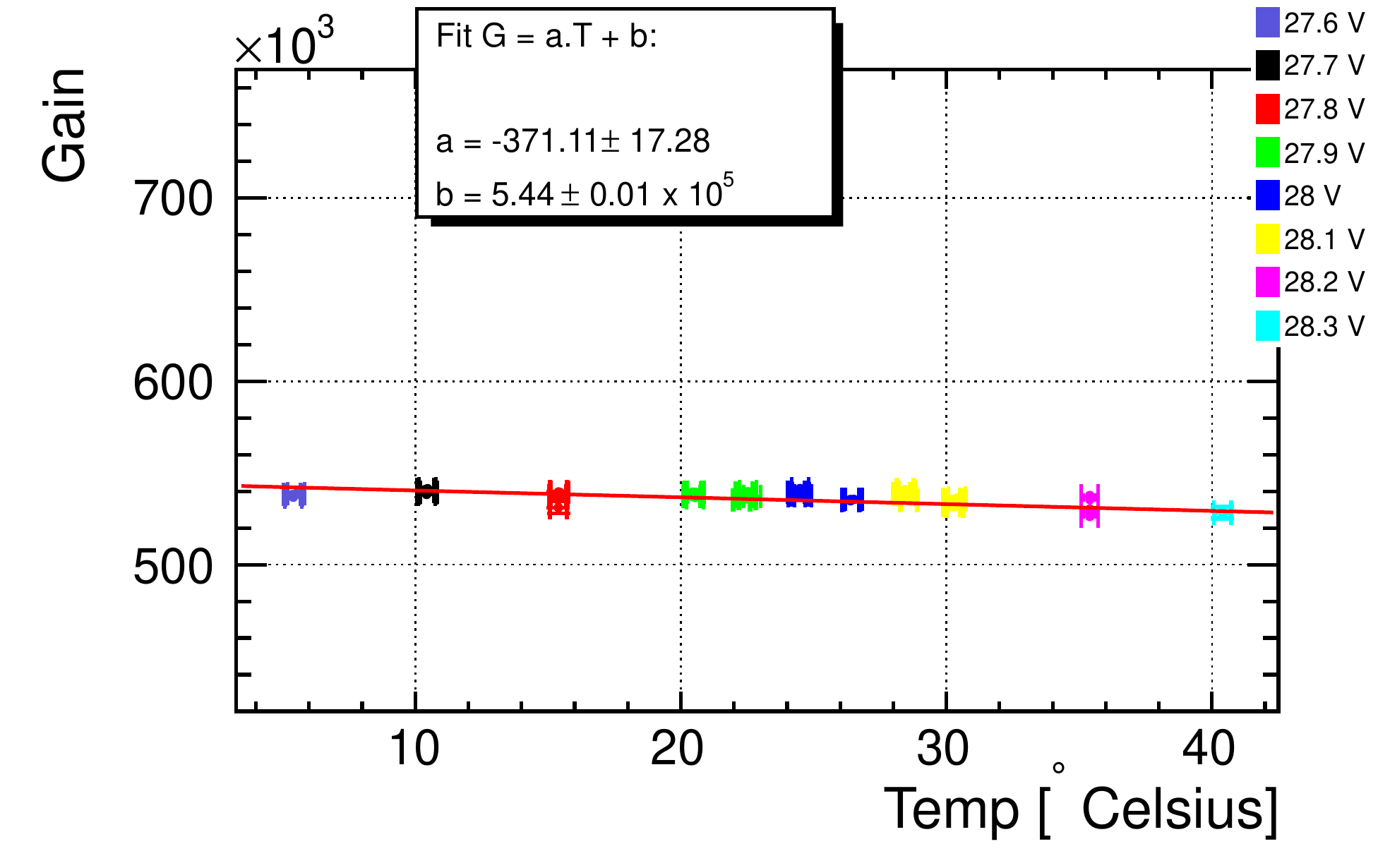}
\caption{Gain versus temperature dependence after bias voltage readjustment for the SiPMs CPTA 857 (top left), CPTA 1677 (top right), KETEK W12 (bottom left), and Hamamatsu 11759 (bottom right). The different colors correspond to different ranges of bias voltage. The insets show the fit results.}
\label{fig:gain-stab}
\end{figure}

\begin{table}[tb]
\vskip 0.2cm
\begin{center}
\begin{tabular}{| l|c|c|c|c|c|}

\hline
\textbf{Detector} &$ \langle$  \textbf{G} $\rangle   [10^5]$  & $ \partial G/ \partial T  [1/^\circ C ] $&   $ \partial G/\partial T ~ [1/^\circ C ] $ & $\sigma_{\Delta G/G}$ & $| \Delta G_T|$ \\
&  \textbf{at $22.5~^\circ  \rm C$}& \textbf{w board} & \textbf{w/o board} &  &  \\
 \hline
$\#857$ & $6.71\pm 0.02 $ & $15\pm 76.$ &-5800 & $\pm 0.31\% $& $ 0.04\%$  \\
$\# 1677$ & $12.8\pm 0.03  $ & $225\pm 106 $ & -14400& $\pm 0.28\% $& $ 0.3\%$  \\
$\#W12$ & $5.44\pm 0.01 $ & $-371\pm 17$ &-38800 & $\pm 0.05\% $& $ 1.2\%$  \\
$\#11759$ & $8.7\pm 0.0 $ & $-859\pm 13$ & -3800& $\pm 0.9\% $& $ 1.7\%$  \\
\hline
\end{tabular}
\end{center}
\caption{Fit results of the gain versus temperature dependence for the offset and slope after gain stabilization, slope before gain stabilization, uncertainty on the gain stability $\sigma_{\Delta G/G}$ and the largest deviation from a constant gain $\Delta G_T$. The latter two observables are determined in the $\rm 10~^\circ  C$ to $\rm 40~^\circ  C$ temperature range.}
\label{tab:gain-fit}
\end{table}

We have tested the gain stabilization method on four SiPMs  including at least one SiPM from each manufacturer.
We adjust the bias voltage continuously using the bias voltage regulator test board. We perform bias voltage readjustments for temperatures between $5~^\circ  \rm C$ and $40~^\circ   \rm C$. At each temperature, we take 16 samples each with 50,000 waveforms. Figure~\ref{fig:dvdt} (right) shows the gain before and after stabilization versus temperature for the CPTA 857 sensor. 

Figures~\ref{fig:gain-stab} depict the results of the gain versus temperature dependence after bias voltage readjustment for SiPMs CPTA 857 (top left), CPTA 1677 (top right), KETEK W12 (bottom left) and Hamamatsu 11759 (bottom right), respectively.  The color code indicates the range of the applied bias voltage, $e.g.$ blue points correspond to the range $ 33.35 < V <33.45~\rm V$.  For SiPM CPTA 857, for example, we perform a linear fit to all data points and obtain an offset of $ (6.71\pm 0.02) \times 10^5$ and a slope of $15.38 \pm 75.87$ indicating that the fit is consistent with a uniform distribution. The deviation from constant gain in the entire temperature range extracted from a linear fit is $|\Delta G_T |<  0.05\%$. Table~\ref{tab:gain-fit} shows the fit results of the gain after and before stabilization, $\sigma_{\Delta G /G}$ and $\Delta G_T$ for all four SiPMs studied. Though the Hamamatsu MPPCs show the largest gain variation and largest $\sigma_{\Delta G/G}$, all tested SiPMs fulfill the requirements of having an uncertainty on the gain stability less than $1\%$.

\section{Determination of the $dV/dT$ dependence}

Gain changes with respect to temperature and bias voltage variations are given by

\begin{equation}
dG(V,T)  = \frac{\partial G(V,T)}{\partial T} \cdot dT + \frac{ \partial G(V,T)}{\partial V} \cdot  dV.
\end{equation}
\noindent
In order to keep the gain constant, we set $ dG=0$  yielding
\begin{equation}
dV/ dT =-\frac{(\partial G/\partial t)}{\partial G/ \partial V}
\end{equation}

\noindent
The ratio $dV/dT$  is obtained from a ratio of two first-order differential equations

\begin{eqnarray}
\frac{\partial G(V,T)}{ \partial T} (V)& = & a+b\cdot V+{\cal O}(V^2), \nonumber \\
\frac{\partial G(V,T)}{\partial V} (T)& =& c+d \cdot T+{\cal O}(T^2),
\end{eqnarray}
\noindent
for coefficients  $a\leq 0$, $b \geq 0, c\geq 0$ and $d\geq 0$.  For the linear case, the analytic solution for $d\neq 0$ is simply

\begin{equation}
 V(T) = \frac{a}{b} +\frac{C}{(c+d\cdot T)^{\frac{b}{d}}},
\end{equation}
\noindent
where $C$ is an integration constant. For $d=0$, the solution is

\begin{equation}
 V(T) = \frac{a}{b} +C \cdot e^{-\frac{b\cdot T}{c}}.
\end{equation}
\noindent
If both $dG/dV$ and $dG/dT$ are constant, the solution is exactly linear
\begin{equation}
 V(T) = \frac{a}{c} \cdot T + C.
\end{equation}
\noindent

\begin{figure}[h]
\centering
\vskip -0.4cm
\includegraphics[width=88mm]{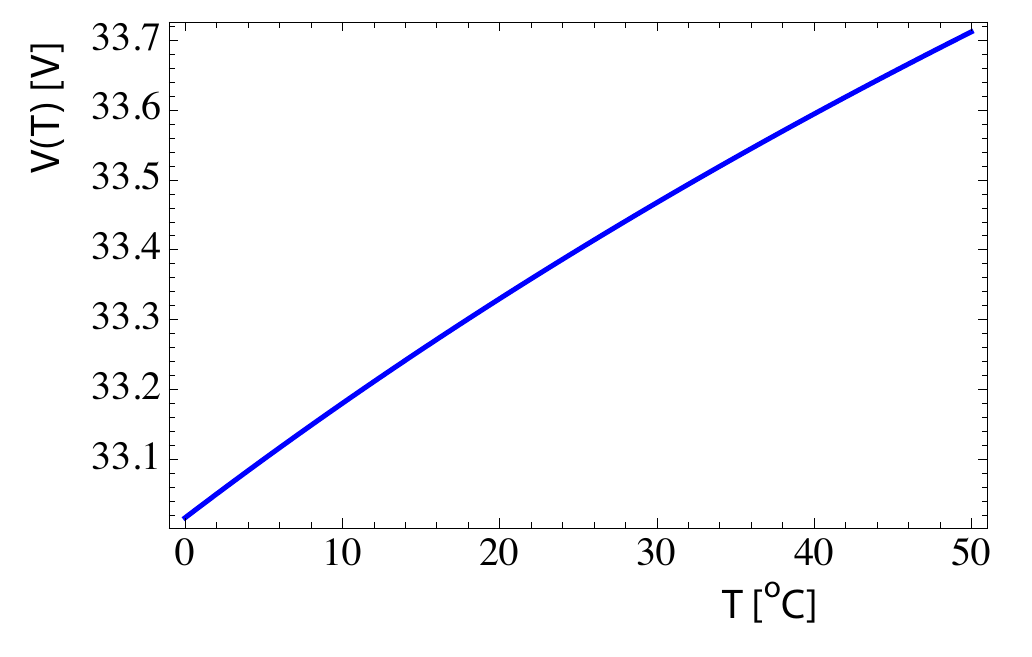}
\caption{The voltage versus temperature dependence determined for SiPM CPTA~857 using a linear model. }
 \label{fig:VT}
\end{figure}

\noindent
As an example, we have calculated the $V(T)$ dependence for SiPM CPTA 857  assuming linear models for  $dG/dT$ and $dG/dV$. Figure~\ref{fig:VT} shows the resulting function $V(T)$.  In the temperature range $20~^\circ  \rm C$ to $30~^\circ   \rm C$, the function is well approximated by a first-order polynomial.

If $dG/dT $ and $dG/dV$ both have a quadratic dependence  

\begin{eqnarray}
\frac{\partial G}{ \partial T} (V)& = & a+b \cdot V+e \cdot V^2, \nonumber \\
\frac{\partial G}{\partial V} (T)& =& c+d \cdot T+f \cdot T^2
\end{eqnarray}
\noindent
with additional coefficients $e$ and $f$, the analytic solution is of the form 
 
\begin{equation}
V(T)=\frac{-b}{2e}+\frac{D1}{2e}
 \tan{  \Bigl (-\frac{D1 \cdot D2} {D3} 
 \arctan{\frac{d+2f\cdot T}
  {D2}- \frac{1}{2}D1\cdot  C}\Bigr )},
\end{equation}

\noindent
where $C$ is again an integration constant and $D1, D2 $ and $D3$ represent 
\begin{eqnarray}
D1&=&\sqrt{-b^2+4 a\cdot e} \nonumber \\
D2&=&\sqrt{-d^2 +4 c \cdot f} \nonumber \\
D3&=&d^2 -4 c \cdot f.
\end{eqnarray}

\section{Conclusion and Outlook}
We have tested 15 SiPMs from three manufacturers of which we selected four to show the performance of the gain stabilization.     
Hamamatsu MPPCs operate at a factor $2.5-3$ higher bias voltage than CPTA and KETEK SiPMs. The $dV/dT$ correction factor for the Hamamatsu MPPCs  ($\simeq -55~\rm mV/^\circ C$) is also about a factor of three higher than that 
 for CPTA and KETEK SiPMs $(15-20~\rm mV/^\circ  C) $.  In the full temperature range $5~^\circ  \rm C~ < T < 40~^\circ   \rm C$, the maximum voltage readjustments are less than $\Delta V=0.7 ~(1.9)~\rm V$ for CPTA and KETEK (Hamamatsu) SiPMs. The deviation from constant gain,  $\Delta G_T$, after bias voltage readjustment is better than  $\pm 0.3\%$ for CPTA, $\pm 1.2\%$ for KETEK and 
 $\pm 1.7\%$  for Hamamatsu SiPMs. Since $\Delta G_T$ depends on $\Delta T$, the deviation from constant gain is reduced by more than a factor of two if we limit the $\Delta T$ range to $20  -30~^\circ   \rm C$. For all four SiPMs, the uncertainty on the gain stability is less than $1 \%$ in the full temperature range ($5~^\circ  \rm C$ to $40~^\circ  \rm C$) as required.
For constant, linear and quadratic  $dG/dT$ and $ dG/dV$ parameterizations, we determined analytical solutions for $V(T)$. 
 
Since the results are encouraging, we have designed a bias voltage regulator board and ordered six prototype boards from an electronic company. We used one board in a recent stabilization test at CERN. We tested five SiPMs in the climate chamber  extending the temperature range from $1^\circ \rm C$  to $50~^\circ  \rm C$. We are presently analyzing the data to check if we need to make any modifications to the bias voltage regulator prototype and order if necessary new bias voltage regulator boards. Furthermore, we plan to implement our correction procedure into the the HBU of the front end electronics of the AHCAL technical prototype at DESY by modifying the power supply and demonstrate that the gain stabilization can be achieved.

%%%%%%%%%%%%%%%%%%%%%%%%%%%%%%%%%%%%%%%%%%%%%%%%%%%%%%%%%%%%%%%%%%%%%%%%%
%%
%%   use this format to include an .eps figure into your paper
%%
%\begin{figure}[htb]
%\centering
%\includegraphics[height=1.5in]{magnet}
%\caption{Plan of the magnet used in the mesmeric studies.}
%\label{fig:magnet}
%\end{figure}
%%%%%%%%%%%%%%%%%%%%%%%%%%%%%%%%%%%%%%%%%%%%%%%%%%%%%%%%%%%%%%%%%%%%%%%%%%%

%%%%%%%%%%%%%%%%%%%%%%%%%%%%%%%%%%%%%%%%%%%%%%%%%%%%%%%%%%%%%%%%%%%%%%%%%
%%
%%   use this format to include a LaTeX table  into your paper
%%
%\begin{table}[t]
%\begin{center}
%\begin{tabular}{l|ccc}  
%Patient &  Initial level($\mu$g/cc) &  w. Magnet &  
%w. Magnet and Sound \\ \hline
% Guglielmo B.  &   0.12     &     0.10      &     0.001  \\
%Ferrando di N. &  0.15     &     0.11      &  $< 0.0005$ \\ \hline
%\end{tabular}
%\caption{Blood cyanide levels for the two patients.}
%\label{tab:blood}
%\end{center}
%\end{table}
%%%%%%%%%%%%%%%%%%%%%%%%%%%%%%%%%%%%%%%%%%%%%%%%%%%%%%%%%%%%%%%%%%%%%%%%%%%

\bigskip
\bigskip

\Acknowledgements
We would like to thank Hamamatsu for providing samples of $\rm 1~mm \times 1~mm$ MPPCs with $\rm 20~\mu m $ and $\rm 15~\mu m $  pixel pitch and KETEK for providing $\rm 3~mm \times 3~mm ~(\rm 2~mm \times 1~mm)$ SiPMs with $\rm 20~(15)~\mu m $ pixel pitch. We further like to thank Lucie Linssen  and Chris Joram for using their photon laboratory. 

%%
%%  bibliographic items can be constructed using the LaTeX format in SPIRES:
%%    see    http://www.slac.stanford.edu/spires/hep/latex.html
%%  SPIRES will also supply the CITATION line information; please include it.
%%

\end{document}